# ADMX Status

**I. Stern[1] on behalf of ADMX**
*University of Florida, Department of Physics*
*Gainesville, FL 32611-8440*
*USA*
*E-mail:* `ianstern@ufl.edu`

Nearly all astrophysical and cosmological data point convincingly to a large component of cold dark matter (CDM) in the Universe. The axion particle, first theorized as a solution to the strong charge-parity problem of quantum chromodynamics, has been established as a prominent CDM candidate. Cosmic observation and particle physics experiments have bracketed the unknown mass of CDM axions between approximately μeV and meV. The Axion Dark Matter eXperim8ent (ADMX) is a direct-detection CDM axion search which has set limits at the KSVZ coupling of the axion to two photons for axion masses between 1.9 and 3.7 μeV. The current upgrades will allow ADMX to detect axions with even the most pessimistic couplings in this mass range. In order to expand the mass reach of the detector, extensive research and development of microwave cavity technologies, tunable microwave SQUID amplifiers, and piezoelectric drives is being conducted. ADMX is projected to explore more than one decade of the allowable mass range with DFSZ coupling sensitivity in the near future. Status of the experiment, current research and development, and projected results are discussed.

Supported by DOE Grants DOE grant DE-SC00098000, DOE grant DE-SC0011665, DE-AC52-07NA27344, DE-AC03-76SF00098, the Heising-Simons Foundation, and the Lawrence Livermore National Laboratory, Fermilab and Pacific Northwest National Laboratory LDRD programs. SQUID development was supported by DOE grant DE-AC02-05CH11231.

---

[1]Speaker

# 1. Introduction

The discovery of dark matter is one of the most compelling explorations in physics [1]. The most widely searched theoretical dark matter particle in is the weakly interacting massive particle, or WIMP. Recent experiments have excluded the existence of the WIMP in much of the plausible parameter space, leaving the science community more perplexed about the composition of dark matter [2]. The axion is another well motivated candidate for dark matter that is both viable, and natural and elegant [3].

## 1.1 Dark Matter Problem

The evidence for dark matter in our universe is abundant. Astrophysical observations that include galactic rotation curves, galaxy cluster dispersions, gravitational lensing, and galaxy cluster mergers strongly suggest the existence of particles that interact almost exclusively through gravitational force [4]. Further indications of this matter are seen across all times scales of cosmology from Big Bang Nucleosynthesis, to cosmic microwave background, to structure formations [5].

While the exact nature of dark matter is yet unknown, select properties of the particle have been determined through indirect observations. Dark matter must have mass and must be long living to cause the rotation curves or gravitational lensing witnessed, for example, and the lack of discovery demonstrates its feeble electromagnetic coupling. Results from the Wilkinson Microwave Anisotropy Probe (WMAP) indicate dark matter is most likely nonrealistic (cold) [6]. Recent measurements from merging galaxy clusters have shown dark matter to have extremely weak self-interaction [7].

## 1.2 Strong CP Problem

The Lagrangian for Quantum Chromodynamics (QCD) contains a charge-parity (CP) symmetry violating term

$$\mathcal{L}_{QCD} = \cdots + \frac{\bar{\theta}g^2}{16\pi^2} G^a_{\mu\nu} \widetilde{G}^{a\mu\nu} + \cdots, \qquad (1)$$

where $G^a_{\mu\nu}$ is the gluon strength tensor and $g$ is a colorless coupling [8]. $\bar{\theta}$ is a consequence of the non-abelian nature of QCD, and is an observable parameter with a value between 0 to $2\pi$ [9]. Without further information, $\bar{\theta}$ would be expected to be of order 1.

As the neutron is composed of quarks, the violation of CP symmetry in QCD should also be observed in the neutron. Indeed, the existence of an electric dipole moment in the neutron would violate CP symmetry, and the strength of the moment could be used to determine the value of $\bar{\theta}$ [10]. To date, no experiments have detected a neutron electric dipole moment, with an upper limit being placed at 3.0 x $10^{-26}$ $e$-cm (at 90% CL) [11]. The resulting upper limit of $\bar{\theta}$ is $\sim 10^{-10}$, which is many orders of magnitude less than predicted. This extremely small upper bound on $\bar{\theta}$ has led to the so-called "strong CP problem."

A solution to the strong CP problem was proposed by Peccei and Quinn. By introducing a new U(1) symmetry, $\bar{\theta}$ becomes a dynamical parameter. Spontaneous breaking of the Peccei-Quinn (PQ) symmetry causes $\bar{\theta}$ to naturally relax to 0 [12]. Weinberg and Wilczek individually showed breaking the PQ symmetry results in the creation of a new pseudo-Nambu-Goldstone boson, dubbed the axion [13].

### 1.3 Axion Dark Matter

Axions naturally and elegantly exhibit all of the predicted properties of dark matter, making them a prominent candidate for dark matter [14]. Further, the axion is predicted to form a Bose-Einstein condensate, which potentially distinguishes it from other cold dark matter candidates through their phase space structure, supported by astronomical observation [15].

The mass of the axion arises from the explicit breaking of PQ symmetry by instanton effects, and is given by

$$m_a \approx 6\,\mu\text{eV}\frac{10^{12}\,\text{GeV}}{f_a}, \qquad (2)$$

where $f_a$ is the axion decay constant, which is proportional to the vacuum expectation value that breaks PQ symmetry [16]. Astrophysical and cosmological observations constrain the mass of axion dark matter, assuming the PQ symmetry is broken post-inflation. The duration of the neutrino burst from supernova SN1987A provided the lower bound on the decay constant of $f_a \gtrsim 10^9$ GeV [17]. The cosmic energy density argument places an upper limit on the decay constant of $f_a \lesssim 10^{12}$ GeV; if the axion energy density were too large, the early universe would have collapsed (overclosure) [18]. From Eq. (2), the allowable mass of axion dark matter is found to be on order between μeV and meV [19].

The electromagnetic interaction of the axion is dictated by the Lagrangian

$$\mathcal{L}_{a\gamma\gamma} = g_{a\gamma\gamma}\, a\, \mathbf{E}\cdot\mathbf{B}, \qquad (3)$$

where $a$ is the axion field, and $\mathbf{E}$ and $\mathbf{B}$ are the electric and magnetic fields of two propagating photons, respectively (see Fig. 1). The coupling constant, $g_{a\gamma\gamma}$, is

$$g_{a\gamma\gamma} = \frac{\alpha g_\gamma}{2\pi f_a}, \qquad (4)$$

where $\alpha$ is the fine structure constant and $g_\gamma$ is a model-dependent constant of order 1 [20]. The lifetime of an axion to decay into two photons is derived from Eq. (3),

$$\tau_{a\gamma\gamma} \approx \left(\frac{10^5\,\text{eV}}{m_a}\right)^5 \text{sec.} \qquad (5)$$

From the limits on $f_a$ above and Eq. (4) and (5), the axion coupling to electromagnetism can be inferred as extremely weak and the lifetime extremely long. For axions with a mass of 1 meV, the coupling constant would be $\sim 10^{-13}$/GeV, and the lifetime would be a $\sim 10^{40}$ seconds ($\sim 10^{22}$ times the age of the universe).

Two theoretical models predict the axion-to-photon coupling for a given mass by defining $g_\gamma$. In the KSVZ (Kim-Shifman-Vainshtein-Zakharov) model, the coupling is [21]

$$g_{a\gamma\gamma}^{\text{KSVZ}} \approx 0.38\frac{m_a}{\text{GeV}^2}. \qquad (6)$$

In the DFSZ (Dine-Fischler-Srednicki-Zhitnitskii) model [22],

$$g_{a\gamma\gamma}^{\text{DFSZ}} \approx 0.14 \frac{m_a}{\text{GeV}^2} \,. \tag{7}$$

## 2. Haloscope Detectors

Because low-mass axions have extremely low decay rates and exceptionally weak interactions with hadronic matter and electromagnetism, they were originally thought to be "invisible" to traditional observational technology [23]. However, Sikivie showed that the decay of dark matter axions is greatly accelerated within a strong static magnetic field through the inverse Primakoff effect [24].

In a static magnetic field, one photon is "replaced" by a virtual photon, while the other maintains the energy of the axion, equal to the rest-mass energy ($m_a c^2$) plus the nonrelativistic kinetic energy. **B** in Eq. (3) is effectively changed to the static magnetic field, **B₀**. The decay rate of the axion increases effectively as $B_0^2$. Figure 1 shows the Feynman diagrams of the axion-photon interaction for the two scenarios.

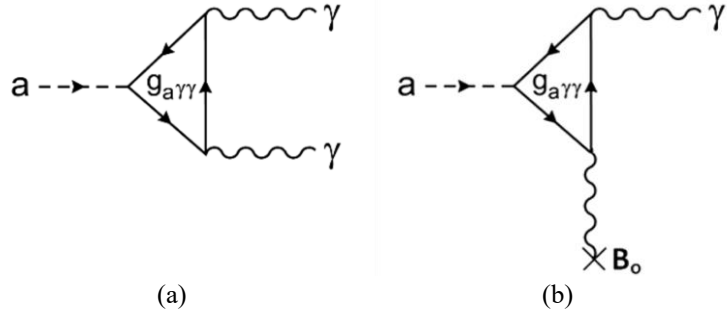

(a)          (b)

**FIGURE 1.** Feynman diagram of axion decay into photons. a) The conversion in vacuum. b) The inverse Primakoff effect in a static magnetic field (**B₀**).

Sikivie proposed an axion dark matter detection scheme based on the Primakoff effect, where a microwave cavity, permeated by a strong magnetic field, is used to resonantly increase the number of photons produced by the axion decay [25]. The axion-photon conversion is enhanced when the resonant frequency $f \approx \frac{m_a c^2}{h}$, where $h$ is Planck's constant. There is also a small correction due to the kinetic energy of the axion, but this is tiny ($\frac{\Delta E}{E} \approx 10^{-6}$) for cold dark matter. Sikivie named this axion dark matter detector a haloscope.

Due to the **E·B₀** term in the Primakoff-decay Lagrangian, a sufficient amount of electric field of the resonant mode must be parallel to the static magnetic field and in-phase throughout the cavity to convert the axion decay into a detectable signal. A normalize form factor is used to quantify the coupling between a cavity mode and the axion decay,

$$C_{mnp} \equiv \frac{\left| \int d^3x \, \mathbf{B_0} \cdot \mathbf{E_{mnp}(x)} \right|^2}{B_0^2 V \int d^3x \left| \mathbf{E_{mnp}(x)} \right|^2} \,. \tag{8}$$

$\mathbf{E_{mnp}(x)}$ is the electric field of a given mode (denoted by the subscripts $m$, $n$, and $p$) within the cavity, and $V$ is the total internal volume of the cavity [26].

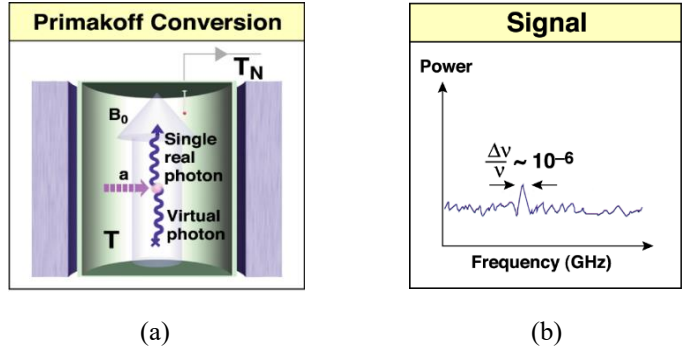

(a) (b)

**FIGURE 2.** a) Graphic representation of a haloscope detector using a solenoid magnet and a cylindrical microwave cavity. b) Anticipated detection signal of a haloscope. The signal can only be observed if the frequency of a mode in the cavity matches the axion energy, and the mode couples to the axion decay.

Figure 2 shows a graphic representation of a haloscope detector and the anticipated axion decay signal. For haloscopes using a cylindrical cavity and a magnetic field pointing parallel to the cylinder axis, Eq. (8) shows only the $TM_{0n0}$ (transverse magnetic) modes will couple to the axion decay. This configuration is the most common for haloscope detectors, with the $TM_{010}$ mode being the strongest coupling mode (thus the best search mode to detect axions). The peak power produced from axion decay for a given mode is

$$P_{mnp} \approx g_{a\gamma\gamma}^2 \frac{\rho_a}{m_a} B_0^2 V C_{mnp} Q_{mnp}, \qquad (9)$$

where $\rho_a$ is the local energy density of the axion field and $Q_{mnp}$ is the quality factor of the cavity for the given mode (assumed to be less than the local energy spread of the axion) [25]. The scan rate determines the integration time of a measurement at one frequency required to meet a specific signal to noise ratio ($s/n$), and is given by [27]

$$\frac{df}{dt} \approx \left(\frac{s}{n}\right)^{-2} \left(\frac{1}{k_B T_n}\right)^2 \frac{g_{a\gamma\gamma}^4 \rho_a^2}{m_a} B_0^4 V^2 C^2 Q_L Q_a. \qquad (10)$$

$T_n$ is the total temperature of the system (physical temperature plus Johnson noise temperature), $k_B$ is Boltzmann's constant, $Q_L$ is the loaded quality factor of the cavity, and $Q_a$ is the effective quality factor of the axion signal. The frequency $f$, form factor $C$, and $Q_L$ are all mode dependent.

## 3. Axion Dark Matter eXperiment (ADMX)

The largest and most sensitive microwave cavity axion search to date is the Axion Dark Matter eXperiment (ADMX). Currently located at the University of Washington, the DOE funded experiment searches for CDM axions with a haloscope detector using a 7.6 Tesla superconducting solenoid and a ~0.15 m³ copper cylindrical microwave cavity. The $TM_{010}$ frequency of the cavity is tuned using two copper rods that are incrementally rotated between scans. The signal measurement is extracted through an antenna critically coupled to the cavity and amplified through a receiver chain. The signal is passed through a crystal filter and mixed down to 35 KHz before being stored in medium-resolution and high-resolution bins for data analysis. The medium-resolution analysis searches for axions signals with a Maxwellian velocity distribution and the high-resolution analysis searches for signals with a fine-structure velocity spread. Figure 3 shows the schematic of the signal processing and a photograph inside a cavity depicting the tuning rods.

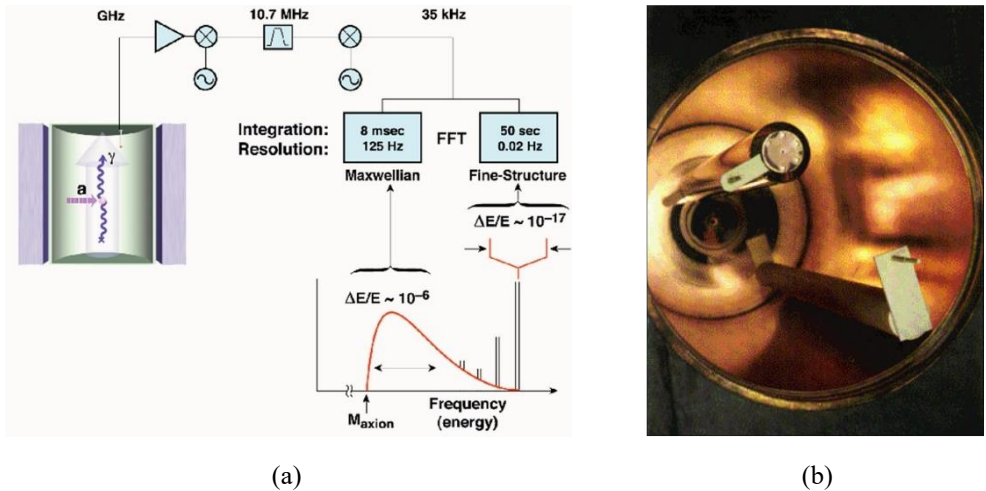

|     |     |
| :-: | :-: |
| (a) | (b) |

**FIGURE 3.** a) Schematic of ADMX receiver chain and data storage bins. b) Photograph displaying the inside of the ADMX microwave cavity. The tuning rods are incrementally rotated about the offset pivot points to adjust the $TM_{010}$ frequency, allowing for scanning across a finite axion mass range.

### 3.1 Previous Searches

The Axion Dark Matter eXperiment began in 1996 at Lawrence Livermore National Laboratory. The initial experiment used two HFET (heterostructure field-effect transistor) amplifiers and was cooled to ~1.3 K, yielding a total temperature of ~6 K [28]. In 2008, ADMX installed a near quantum-limited SQUID (superconducting quantum interference device) amplifier, reducing the total temperature to ~3 K. In order to operate at microwave frequencies, the input coil was replaced by a resonant input microstrip. A bucking magnet was also added to the experiment to counter the magnetic field from the main magnet locally at the SQUID [29]. Figure 4 shows a graphic illustration of the ADMX experiment with the SQUID amplifier and bucking magnet, a graphic illustration and a photograph of a SQUID amplifier, and the magnet.

ADMX has conducted searches for a number of years, successfully excluding axions with KSVZ coupling strength in the mass range of 1.9–3.7 µeV (460–890 MHz) [30]. Figure 5 shows the published limits for the experiment.

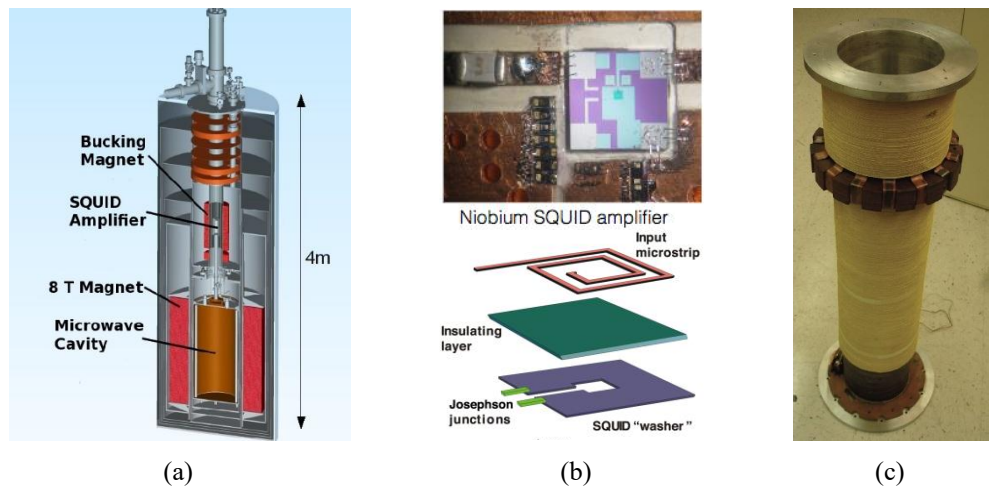

|     |     |     |
| :-: | :-: | :-: |
| (a) | (b) | (c) |

**FIGURE 4.** a) Graphic illustration of ADMX experiment. b) Top: photograph of a SQUID amplifier. Bottom: graphic illustration of a SQUID amplifier. c) Superconducting solenoid "bucking" magnet.

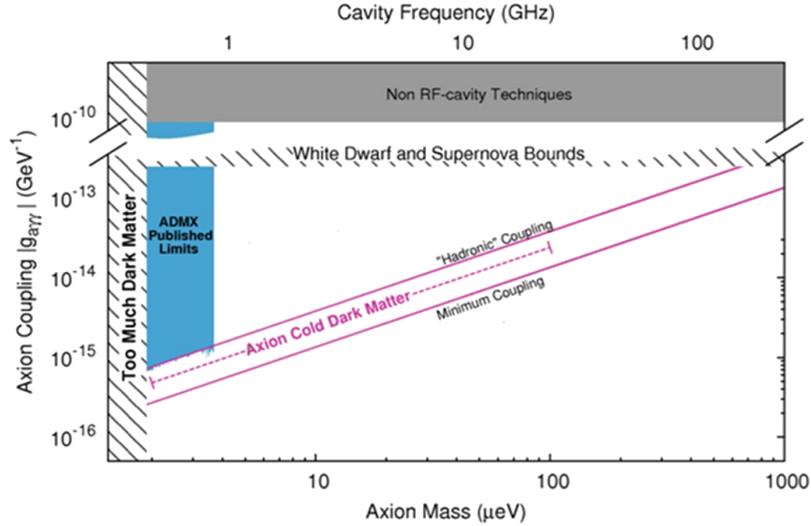

**FIGURE 5.** Axion exclusion plot showing published results of ADMX. The KSVZ coupling is labeled "Hadronic" and the DFSZ coupling is labeled "Minimum".

### 3.2 Phase IIa/Gen 2 Upgrades

ADMX has recently installed several enhancements to the detector to improve signal sensitivity and expand the axion mass search range. The Phase IIa/Gen 2 upgrades included purchasing and commissioning a high-cooling power dilution refrigerator, developing and integrating a second, smaller, tunable cavity ("Sidecar"), and incorporating two additional search channels. ADMX has completed engineering scans with the upgraded system. The improvements will allow the detection of axions below DFSZ coupling and searches at new masses.

The dil. fridge maintains 8 mK base temperature and demonstrated the design specified 800 μW cooling power at 100 mK. The dry system is mounted directly above the cavity and is thermally linked to the SQUID amplifier to cool both components. During commissioning, the cavity was cooled to 200 mK. Further modifications are being made to the cryogenic system to reduce the cavity temperature below 150 mK for future searches. Figure 6(a) shows the dilution refrigerator installed on the ADMX cavity.

The Sidecar cavity is approximately $1/10^{th}$ the size ($1/1000^{th}$ the volume) of the ADMX main cavity, and is thermally mounted to the top of the main cavity well within the magnetic field of the solenoid. The detector uses a single copper tuning rod to search for axion dark matter with mass of 14-25 μeV (3.5-6.0 GHz). Due to the greatly reduced volume, the Sidecar is not sensitive enough to detect axions at the KSVZ coupling (with predicted local axion density); the experiment serves primarily as a pathfinder for piezoelectric motors and research for higher-mass cavities (see section 3.3). Figure 6(b) shows the Sidecar mounted on top of the ADMX cavity.

Two additional receiver channels were integrated into the ADMX electronics. One channel is dedicated to the Sidecar detector and the other is used to probe the $TM_{020}$ mode of the main detector. The $TM_{020}$ channel will search for axions with mass of 4.5-6.0 μeV (1.1-1.5 GHz). The cavity coupling to the axion decay ($C$) of the $TM_{020}$ mode is approximately ¼ the coupling of the $TM_{010}$ mode. With the increased cooling power from the dil. fridge, the $TM_{020}$ search channel will be able to detect axions below KSVZ coupling. To reduce data analysis time, the signal in all channels is now directly digitized and digitally down-sampled prior to storage.

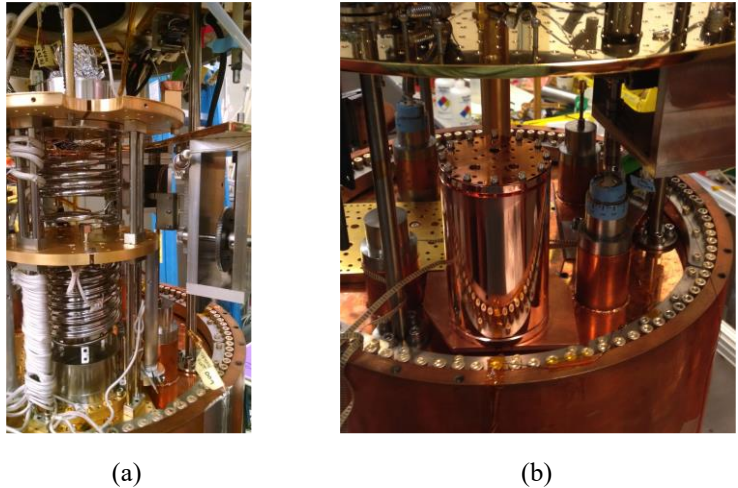

(a) (b)

**FIGURE 6.** a) Dilution refrigerator installed on top of the ADMX main cavity. b) Sidecar cavity mounted on top of the ADMX main cavity.

### 3.3 Gen 2 Research and Development

The Gen 2 ADMX experiment is scheduled to search for axion dark matter in the mass range of 2-40 µeV (0.5-10 GHz) down to DFSZ coupling. To achieve this goal, several multiyear R&D projects have been established. The Cavity Working Group is undertaking long-term studies to design and fabricate the next generation microwave cavities for higher-mass searches. Piezoelectric motor drives are being tested to reduce heat loads, and tunable high-frequency SQUID amplifiers and Josephson parametric amplifiers (JPA) are being developed to increase gain. The program has also initiated studies and cost analyses of high-field magnets.

In order to achieve more than 1 order of magnitude search capabilities, a tiered frequency method will be implemented. The tiers are divided into the frequency ranges 0.5-1 GHz, 1-2 GHz, 2-4 GHz, 4-6 GHz, 6-8 GHz, and 8-10 GHz. The first frequency range will be achieved by the current ADMX cavity. The second frequency tier will utilize 4 cavities that are frequency locked using the Pound stabilizer system [31]. The 2-4 GHz and 4-6 GHz tiers will use ~16 and ~32 frequency locked cavities, respectively. The 6-8 GHz range will require a photonic band-gap cavity, a complex system of multiple tuning posts that move in concert to maintain symmetry to maximize coupling, *C*. The final tier configuration is still in the early stages of research.

To obtain DFSZ sensitivity at higher masses, sophisticated manufacturing practices are being implemented to minimize mode mixing and symmetry breaking, two anomalies that greatly reduce *C* within the search range, as well as maximize the quality factor of the cavities. Superconducting hybrid cavities are also being investigated. Hybrid cavities would utilize superconducting thin-film coats on the vertical surfaces to increase the *Q* of the detector ~10 fold. Additionally, in situ mode identification techniques and field characterization methods are being tested to improve reliability of scans. Figure 7(a-b) illustrates the 4-cavity design and shows a graphic illustration of a photonic band-gap cavity along with a photograph inside a prototype.

ADMX currently uses worm gears to meet the fine motion control requirements of the experiment. The gears produce significant heat and such heat will be prohibitive with a high number of cavities. Thus, piezoelectric motor drives are being tested on the Sidecar for future searches. Figure 7(c) shows a photograph of the motor drive prototype.

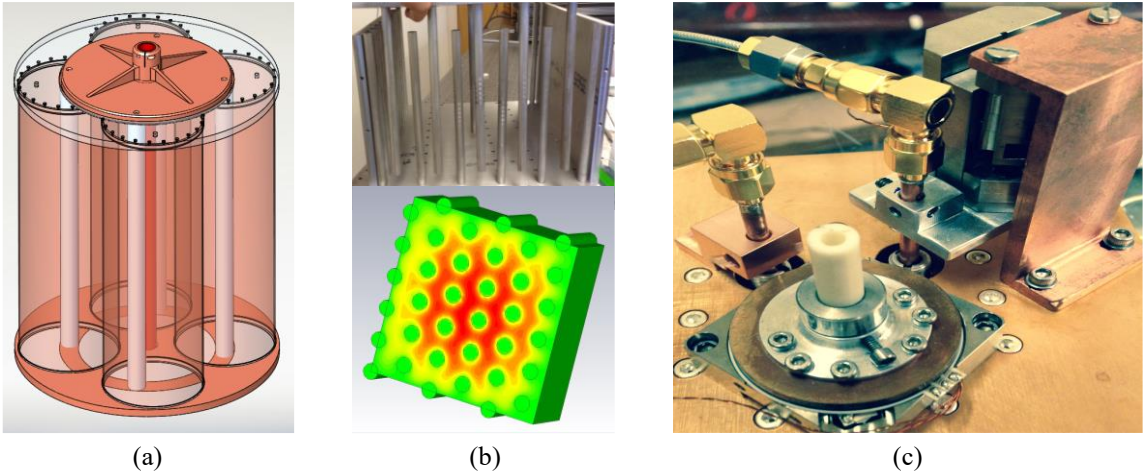

**FIGURE 7.** a) Graphic illustration of the 1-2 GHz cavity design. The Pound stabilizer will be used to frequency lock the cavities. b) Top: photo of a prototype photonic band-gap being assembled. Bottom: results of a simulation showing the electric field strength. Symmetry breaking will cause significant localization of the electric field and a reduction in *C*. c) Piezoelectric motor drive prototype.

Microwave SQUID amplifiers using varactor tuning have been proven. The amplifiers enable the detector to maintain maximum gain throughout the search. Further research is being conducted to maintain near quantum-limited noise across the tuning range. JPA will need to be utilized at higher frequency levels and are being procured by the program. Figure 8 shows the results of the varactor tuning of a microwave SQUID amplifier.

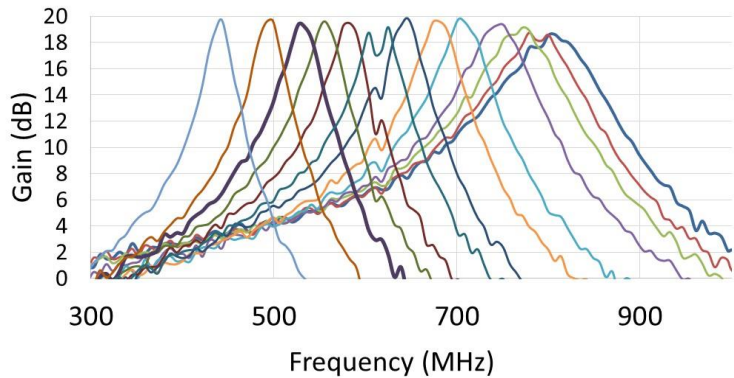

**FIGURE 8.** Microwave SQUID amplifier demonstrating nearly an octave of tuning at ~20 dB gain.

The current R&D is expected to enable ADMX to meet the Gen 2 goal of searching up to 40 µeV (10 GHz) by the year 2022. Figure 9 shows the projected sensitivity of ADMX Gen 2, incorporating all detector enhancements, depicting the frequency tiers separate by color.

## 4. Conclusion

The evidence for cold dark matter in our universe is abundant. The "invisible" axion has been shown to be a prominent cold dark matter candidate within the mass range of approximately µeV and meV. The Axion Dark Matter eXperiment is the only axion detector to demonstrate KSVZ sensitivity and has ruled out axions with KSVZ coupling strength in the mass range of 1.9–3.7 µeV (460–890 MHz). Phase IIa and Gen 2 upgrades to the ADMX detector are complete and have been proven through full-system engineering test runs. The current experiment configuration will be the first axion dark matter detector to search with DFSZ sensitivity. Scientific testing is scheduled to start in 2016, and will probe new masses using the $TM_{020}$ channel and the Sidecar detector. ADMX is aggressively researching new microwave cavity technologies, as well as tunable microwave SQUID amplifiers, JPA, and piezoelectric motor drives, to expand the detectors search capabilities up to 40 μeV (10 GHz) with DFSZ sensitivity by 2022.

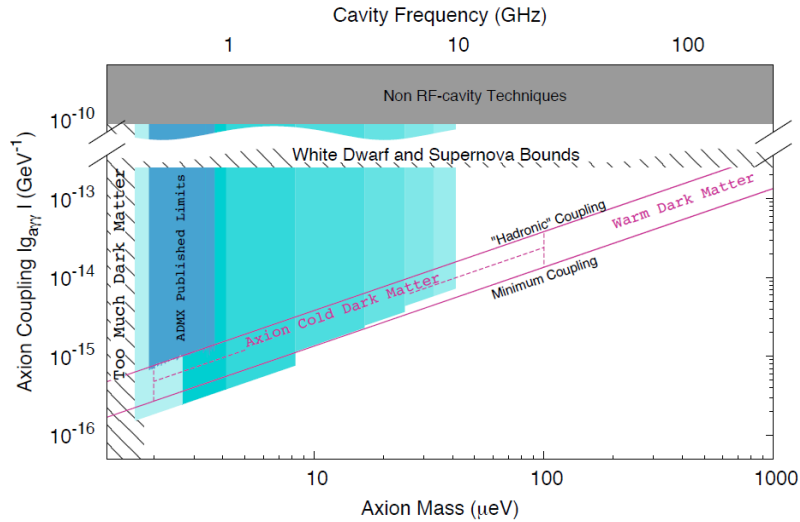

**FIGURE 9.** Projected sensitivity of ADMX Gen 2. Searches are scheduled to be complete by 2022.